%% file: paper.tex
\documentclass[conference]{IEEEtran}
\IEEEoverridecommandlockouts
\usepackage{cite}
\usepackage{amsmath,amssymb,amsfonts}
\usepackage{algorithmic}
\usepackage[table,xcdraw]{xcolor}
\usepackage{graphicx}
\usepackage{textcomp}
\usepackage[colorinlistoftodos,prependcaption]{todonotes}
\presetkeys{todonotes}{inline}{}
\usepackage{pgfplots}
\usepackage{subcaption}
\usepackage{listings}
\usepackage{url}

\usepackage{xcolor}
\def\BibTeX{{\rm B\kern-.05em{\sc i\kern-.025em b}\kern-.08em
    T\kern-.1667em\lower.7ex\hbox{E}\kern-.125emX}}

\newcommand\copyrighttext{%
  \footnotesize \textcopyright 2020 IEEE. Personal use of this material is permitted. Permission from IEEE must be
obtained for all other uses, in any current or future media, including
reprinting/republishing this material for advertising or promotional purposes, creating new
collective works, for resale or redistribution to servers or lists, or reuse of any copyrighted
component of this work in other works.}
\newcommand\copyrightnotice{%
\begin{tikzpicture}[remember picture,overlay]
\node[anchor=south,yshift=10pt] at (current page.south) {\fbox{\parbox{\dimexpr\textwidth-\fboxsep-\fboxrule\relax}{\copyrighttext}}};
\end{tikzpicture}%
}

\begin{document}

\title{Towards AIOps in Edge Computing Environments\\
}

\author{
    \IEEEauthorblockN{Soeren Becker\IEEEauthorrefmark{1}, Florian Schmidt\IEEEauthorrefmark{1}, Anton Gulenko\IEEEauthorrefmark{1}, Alexander Acker\IEEEauthorrefmark{1}, Odej Kao\IEEEauthorrefmark{1}}
    \IEEEauthorblockA{\IEEEauthorrefmark{1}\textit{Complex and Distributed IT-Systems} \\ \textit{TU Berlin}\\ Berlin, Germany
    \\\{firstname\}.\{lastname\}@tu-berlin.de}
}

%\author{\IEEEauthorblockN{ S\"oren Becker}
%\IEEEauthorblockA{\textit{Complex and Distributed IT-Systems} \\
%\textit{TU Berlin}\\
%Berlin, Germany \\
%soeren.becker@tu-berlin.de}
%\and
%\IEEEauthorblockN{ Florian Schmidt}
%\IEEEauthorblockA{\textit{Complex and Distributed IT-Systems} \\
%\textit{TU Berlin}\\
%Berlin, Germany \\
%florian.schmidt@tu-berlin.de}
%\and
%\IEEEauthorblockN{ Anton Gulenko}
%\IEEEauthorblockA{\textit{Complex and Distributed IT-Systems} \\
%\textit{TU Berlin}\\
%Berlin, Germany \\
%anton.gulenko@tu-berlin.de}
%\and
%\IEEEauthorblockN{ Alexander Acker}
%\IEEEauthorblockA{\textit{Complex and Distributed IT-Systems} \\
%\textit{TU Berlin}\\
%Berlin, Germany \\
%alexander.acker@tu-berlin.de}
%\and
%\IEEEauthorblockN{Odej Kao}
%\IEEEauthorblockA{\textit{Complex and Distributed IT-Systems} \\
%\textit{TU Berlin}\\
%Berlin, Germany \\
%odej.kao@tu-berlin.de}
%}

\maketitle
\copyrightnotice
\begin{abstract}
Edge computing was introduced as a technical enabler for the demanding requirements of new network technologies like 5G. It aims to overcome challenges related to centralized cloud computing environments by distributing computational resources to the edge of the network towards the customers. The complexity of the emerging infrastructures increases significantly, together with the ramifications of outages on critical use cases such as self-driving cars or health care. Artificial Intelligence for IT Operations (AIOps) aims to support human operators in managing complex infrastructures by using machine learning methods. This paper describes the system design of an AIOps platform which is applicable in heterogeneous, distributed environments. The overhead of a high-frequency monitoring solution on edge devices is evaluated and performance experiments regarding the applicability of three anomaly detection algorithms on edge devices are conducted. The results show, that it is feasible to collect metrics with a high frequency and simultaneously run specific anomaly detection algorithms directly on edge devices with a reasonable overhead on the resource utilization.
\end{abstract}

%self-healing, 

\input{01-introduction}

\input{06-related-work}

\input{02-background}

\input{03-platform.tex}

\input{04-experiments.tex}

\input{07-summary}

\bibliographystyle{IEEEtran}
\bibliography{paper}

\end{document}

%% file: 01-introduction.tex
\section{Introduction}
Artificial Intelligence for IT Operations (AIOps) describes the process of maintaining and operating large IT infrastructures using AI-supported methods and tools on
different levels, e.g. automated anomaly detection and root cause analysis, remediation and optimization, as well as fully automated initiation of self-stabilizing
activities. The automation is mandatory to handle the increasing system complexity due to the virtualization and software-defined everything trends, combined with the
significant increase in numbers of servers, devices, sensors. %, and applications interacting with the infrastructure. 
Service providers are aware of the 
need for always-on, dependable services and already introduced additional intelligence to the IT-ecosystem, e.g. by employing network and site reliability engineers, 
by deploying automated tools for 24/7 monitoring, %and AIOps platforms for load balancing, 
and capacity planning.%, resource utilization, storage management, and threat detection.
%This decision-making in combination with advanced virtualization techniques allow a significant progress regarding reliability, serviceability, and 
%availability: checkpointing, migration, re-starts, scale-out/scale-down, re-routing, resource reservations, and other operations are much more flexible and easier
%to deploy, even independent from the underlying hardware infrastructure and not limited to a certain geophysical location. 

Rapidly decreasing the reaction time by system administrators is still highly encouraged and necessary due to performance problems (tuning), 
to component/system failures (outages, degraded performance), or due to security incidents. Quick latency-bounded responses are demanded in multiple scenarios ranging from smart cities via automated manufacturing to autonomous driving \cite{Shi2016}. %  of 
%latency-bounded responses in multiple scenarios ranging from smart cities via automated manufacturing to autonomous driving . 
In many situations, a response within 
pre-defined latency limits is not a soft requirement but can have a serious impact on the system functionality in our everyday life. A lot of efforts
were spent in the last decade to move the actors and sensors closer to the source of data and reduce or even eliminate the uncertainty of the network performance, 
leading to the rise of edge and fog computing. 
The management of edge and fog computing environments exposes – next to the obvious increase of complexity through quantity
of devices – to a number of additional drawbacks. The devices are typically physically located outside of data centers. Consequently, the devices are vulnerable to damaging actions, intentionally 
(theft, damage, jamming, etc.) and naturally (weather influence, aging, interferences, etc.). Moreover, the spatial distribution of such devices prevents the application
of standard data center procedures related to redundancy, access control, or maintenance. These drawbacks create a paradox situation: a vulnerable infrastructure has a 
decisive impact on our everyday life, as it delivers crucial data for many autonomous and smart-X processes. 
Thus, it is important to enable the continuous analysis of all relevant measurements to ensure correct minimal functionality of edge and fog devices.
%all available measures for 
%a continuous and correct minimal functionality of involved edge and fog devices.

During previous work we have developed a self-healing pipeline which was extensively tested in cloud computing environments \cite{Gulenko2016b}. It consists of detecting autonomously anomalies, find the root cause, and plan and execute appropriate actions to resolve or mitigate the occurred problem. However, the requirements for edge and fog computing demand modifications to ensure the feasibility of the analysis pipeline in highly distributed environments.

We aim contributing to this goal by addressing:% the following relevant aspects:

%\begin{itemize}
%    \item Continuous monitoring of all integrated devices and in-place analysis - in proximity to the data sources - to detect anomalies and upcoming faults
%    \item Self-healing methods to remediate the anomalies and ensure an uninterrupted functionality of the devices itself or a group of related devices. 
%\end{itemize}

\begin{itemize}
    \item Support heterogeneous CPU architectures to enable in-place analysis on edge devices of the self-healing pipeline
    \item Show the feasibility to apply monitoring and AI models on edge devices with limited compute resources
   \item Environment-aware placement of analysis steps in the infrastructure to prevent overloading of lightweight edge devices and to comply with requirements of AI models on the hardware.
\end{itemize}

This paper describes the AIOPs platform ZerOps4E which combines monitoring and AI-based methods to 
learn the normal behavior of the edge devices and to detect deviations in this behavior indicating anomalies. We assume that edge devices are capable of running
at least a docker container (for example Raspberry Pis). Thus, the ZerOps4E platform can be executed directly on the edge. This allows a local execution of 
all monitoring and at least anomaly detection services, which reduces the requirements on network bandwidth and provides a foundation for a low-latency processing.
With a series of experiments, the performance and resource overhead of the monitoring agent and different anomaly detection algorithms on edge devices are tested to evaluate the feasibility of running those framework components directly at the edge.

The remainder of the paper is organized as follows: Section II outlines the related work whereas Section III describes the background of self-healing analysis, which are applied in ZerOps4E. Section IV proposes the overall architecture of ZerOps4E and introduces the main framework components for data collection and self-healing analysis. In Section V the performance experiments are illustrated and evaluated, while Section VI concludes the main findings.

%% file: 06-related-work.tex
\section{Related Work}
Anomaly detection, predictive maintenance and AIOps in general are important topics especially for highly distributed environments such as edge computing infrastructures \cite{Yu2017}. 
Bose et al. \cite{Bose2019} for instance are applying anomaly detecion methods in the predictive maintenance context: For the lifecycle of edge devices they start with low accuracy models and only switch to higher accuracy models when anomalies are detected to decrease the energy consumption.

Another approach \cite{Schneible2017} is leveraging federated learning to train models directly on edge devices and aggregate the resulting weights by a central component. The authors are using deep learning neural networks to identify anomalies and further exchange model data between edge nodes and the central component when an internet connection is available. They show that federated models can produce similar results to non-distributed models for several data sets.

Soualhia et al. \cite{Soualhia2019} are proposing a framework similar to ours: In contrast, they are mainly applying supervised machine learning techniques to detect and predict faults whereas we utilize unsupervised methods. Components for pre-processing, fault detection and fault prediction are provided and evaluated on a simulated edge computing environment while injecting synthetic faults. Their approach shows promising results for non-fatal CPU and HDD overload anomalies. The authors further plan to test their components and models on resource-constrained environments. 

Furthermore, anomaly detection is also often applied to detect attacks or malicious behaviour, for example in \cite{Kozik2018}: Kozik et al. propose a platform to classify anomalous network traffic on edge devices by using extreme learning machine classifiers trained in the cloud. Another approach is introduced in \cite{lyu2017fog} where the authors are deploying anomaly detection algorithms in the fog layer instead of directly on the edge devices: The proposed \textit{Fog-Empowered} mechanism is able to improve the processing delay and energy consumption compared to other centralized and distributed methods by using  hyperellipsoidal clustering. 

Not only the scientific community has recognized the importance of AIOps, several industrial player are also working on solutions:

The vision of the Fixstream AIOps+ platform\footnote{https://fixstream.com} is to predict any issues that can impact business
applications and to automatically remediate these issues before they result in business outages.
They combine business transactions, application and
infrastructure issues in a single root cause analysis and thus aim optimize IT resources to reduce the
infrastructure costs.

Moogsoft\footnote{https://www.moogsoft.com/} offers a closed source AIOps platform which uses different machine learning
algorithms to predict anomalies on a given data stream. The main purpose is recognizing root causes of failures within large infrastructures. 
Based on supervised training, Moogsoft trains a machine learning model predicting the root cause for
similar failures in the future.  The
user can then choose a remediation that is afterwards stored in a database to be recommended
for future problems.

%Another company offering a AIOps solution is Grok\footnote{https://grokstream.com/}. Grok analyses streaming telemetry
%data to predict abnormal activity in various environments. When anomalies are predicted or
%found an event is triggered which in turn can be used to automatically resolve the issue. Grok
%states that they are able to differentiate seasonal changes in data from anomalies.

%% file: 02-background.tex
\section{Background Self-healing Analysis}

For data collection, we utilize the open-source project bitflow-collector\footnote{\url{https://github.com/bitflow-stream/go-bitflow-collector}}. The monitoring service is able to collect resource metrics from several different hardware interfaces.
%, the core operation and virtualization processes. The collector is written in Golang\footnote{\url{https://golang.org/}} and is able to be configured in its monitoring rates in arbitrary high frequencies while promising to keep the resource overhead low. 
The bitflow-collector monitors metrics for VMs i.e. through libvirt\footnote{\url{https://libvirt.org/}} and physical servers by parsing the proc-file system\footnote{\url{https://www.tldp.org/LDP/Linux-Filesystem-Hierarchy/html/proc.html}}.
The monitoring service can be configured to collect different monitoring rates in arbitrary high frequencies while promising to keep the resource overhead low.

Gulenko et al. \cite{Gulenko2016b} described an analysis pipeline and architecture to provide self-healing to cloud systems. The system is build upon the open-source stream processing framework Bitflow\footnote{https://github.com/bitflow-stream/} in order to enable the sending of monitoring data as well as analysis results between machine learning algorithms.

The analysis pipeline consists of the following steps:
%\begin{itemize}
    %\item 
    
    \textbf{Anomaly detection}: For each monitored component an individual machine learning model is deployed. The model applies an unsupervised online algorithm, which is capable of detecting autonomously abnormal behavior of resource patterns. In case of abnormality, the root cause analysis is alerted and the state of all components are forwarded.% of the monitored component is forwarded to the root cause analysis. 
    
    %\item 
    \textbf{Root cause analysis}: The root cause analysis investigates the actual source of the underlying problem, i.e. the component causing the anomaly. It requires knowledge about the infrastructure's horizontal and vertical dependencies.
    
    %\item 
    \textbf{Remediation engine} (decision engine): The engine aggregates the analysis results and combines the decisions to schedule and manage the execution of appropriate actions. This service requires the existence of a catalogue of remediation actions, which are expected to be provided by experts. Furthermore, a mapping is required to match the anomalies and component of cause to the action.
%\end{itemize}

%\begin{figure*}[h]
%    \centering
%    \includegraphics[width=0.9\textwidth]{images/zerops_arch.jpg}
%    \caption{Architecture of an OpenStack cloud infrastructure with a %self-stabilizing pipeline.}
%    \label{zerops_arch}
%\end{figure*}

This architecture provides a flexible structure to apply several different machine learning approaches to the analysis pipeline, making it flexible to investigate different approaches in this field. 
The proposed architecture lacks of the flexibility of placing analysis steps on physically distributed compute devices and only applies streaming connections between analysis steps, causing highly intensive network traffic. In order to overcome these problems, we propose an adapted platform, which is specialized on the needs and requirements of highly distributed and heterogeneous infrastructures. Additionally, we describe potential candidates of machine learning techniques to be placed within the infrastructure.

%% file: 03-platform.tex
\section{ZerOps4E}
The ZerOps4E AIOps platform leverages the lightweight Bitflow stream processing framework to enable in-place analysis of 
data streams\cite{gulenko2020bitflow}. The term in-places refers to the processing of data streams as close as possible to their sources.
%: Since the edge computing paradigm
%typically involves an environment consisting of multiple physically apart sites which are not located in the same data center but rather spatial distributed
%(e.g. across a city in a Smart City use case), it is unfeasible to stream all metrics to a central analysis component. 
The network can congest due to 
multiple simultaneous data streams, which could further lead to increased duration and fluctuation of data processing steps. We assume, that in-place analysis
decreases the network load and is therefore in line with the strong requirements of edge computing.

We are using Kubernetes as underlying infrastructure for the platform since containerization is an effective alternative to traditional virtualization on lightweight edge devices \cite{Bellavista2017a}. 
%Due to the low overhead of containerization it is often used as an orchestration framework in fog and edge computing environments \cite{Bellavista2017a,Hoque2017,Pahl2016,Santoro2017, Ismail2016} which further motivates the integration of the AIOps platform in Kubernetes.

\subsection{Data Collection}
We are utilizing the bitflow-collector as it promisses low resource consumption for monitoring, which is needed for low-powered edge devices. 
%investigated further
%\footnote{\url{https://github.com/bitflow-stream/go-bitflow-collector}} 
%to monitor data from several hardware components and the core operation and virtualization processes. The collector is written in Golang\footnote{\url{https://golang.org/}} and is able collect metrics in arbitrary high frequencies while keeping the resource overhead low. 
%We evaluated the overhead of the collector with different frequencies while running on edge devices and show the results in section (TODO). 
The data is converted into a binary transport format and provided as a data stream which can be consumed by the data analysis steps. In addition, the collector was extended with a plugin to automatically create a data source object for the monitored node in the Kubernetes cluster.

\subsection{Data Analysis}
The self-healing analysis pipeline consists of several algorithms chained together in order to detect problems on a component level, while aggregating the results into a higher-level decision engine, which plans and executes appropriate actions.
The data analysis pipeline is split into three parts:
\begin{itemize}
    \item \textbf{Decentralized analysis}: Processing and filtering of data is applied in a first place near or directly on the compute device of the data source. An edge case is the applicability of applying anomaly detection as decentralized component as it might consume too much resources on the low-powered edge devices. From its concept, the anomaly detection should be placed next to the data source, whenever the compute capabilities are available. Otherwise, the anomaly detection has to be moved to near by components or to the cloud. This edge case is increasingly interesting as it provides benefits when placing the detection directly on the edge device as network traffic can be saved. Therefore, we investigate the feasibility of this key technique in the evaluation part.
    \item \textbf{Centralized analysis}: This contrasting concept consists of analysis pipelines, which need to be executed centralized, for instance in a traditional cloud service. The decision engine, which orchestrates the execution of remediation actions on a global level is such kind of analysis step, which requires the knowledge of the whole infrastructure. Consequently, such algorithms require the existence of global information and due to their centralized deployment, intensive computational resources can be assured. In contrast, the latency might increase due to the physical distances between monitored edge devices and the cloud servers \cite{Pfandzelter2019}.
    \item \textbf{Partly-centralized analysis}: Besides the extreme cases of  decentralized and centralized appliance of analysis steps, further cases exist in between. For example, root cause analysis can be applied with respect to a network slice or server rack instead of a global level.
\end{itemize}

In order to ensure the transportation of data between the placed algorithms, we leverage the Bitflow framework for high throughput data streams, but on higher-level computations, we utilize message queues. Stream processing can provide near real-time insights \cite{Pfandzelter2019} whereas aggregated events are exchanged using the event bus.
ZerOps4E implements RabbitMQ\footnote{\url{https://www.rabbitmq.com/}} as event bus to transfer events - i.e. results of the decision engine -  rather than streams.

%The Bitflow framework supports a light-weight domain specific language (DSL) for definition of data processing pipelines.
%This scripting language allows the flexible construction of data analysis steps from a toolbox of predefined components like shown in listing 1. Bitflow already provides
%several preprocessing steps such as data normalization or filtering. These preprocessing steps can be combined in many ways to shape the incoming data to 
%best fit the actual machine learning algorithm. In addition to preprocessing steps Bitflow also provides several anomaly detection algorthms which can
%be used to detect anomalous situtations \cite{Gulenko2016c}.    

%\begin{lstlisting}[caption={Bitflow Script example which uses a Data Stream as input, standardizes the values, detect anomalies and finally stores labeled results in a csv file }]
%    Data stream input -> Standardize() 
%                      -> DetectAnomalies()
%                      -> output.csv
%\end{lstlisting}

%As preprocessing we apply min-max scaling to standardize the different input dimensions to the same scaling. Most min-max values can be automatically inferred from the underlying hardware like RAM, disc space etc., while further min-max values like network traffic are assumed to be tested beforehand through stress tests.

The implementation of the anomaly detection algorithms follow the principle of Identity Function and Threshold Model \cite{Schmidt2018} to automatically adjust the anomaly detection model to the evolving data stream in an unsupervised manner.
We integrated  Long-short term memory (LSTM) \cite{Schmidt2018}, BIRCH \cite{Zhang1996a} and Autoregressive integrated moving average (ARIMA) \cite{Schmidt2018a} as reconstruction functions, while applying exponential moving average as dynamic threshold model.
As anomalies might propagate between monitored components, we apply a time-based root cause analysis. Thus, components are stated as root cause components, where anomalies can be detected earlier. Of course, this technique is limited in complex systems, we therefore recommend the survey by Solé et al. \cite{Sole2017a} for choosing advanced approaches.

Lastly, the decision engine gathers events of root cause components with additional information about the anomalies and applies density grid pattern matching \cite{acker2018online} to recommend appropriate remediation action selection. These actions are assumed to be provided beforehand by experts, but can be extended over time through reinforcement learning. Actions are for example the migration of a virtual machine or reconfiguration of a particular service.

\subsection{Operator Component}
We extended the bitflow-k8s-operator\footnote{https://github.com/bitflow-stream/bitflow-k8s-operator}, which orchestrates the deployment of the distributed deployment of the analysis pipeline. It is responsible for parsing the infrastructure dependency model and
scheduling analysis pipelines in the infrastructure. It leverages the Kubernetes concept of custom resource definitions (CRDs) to introduce
two custom objects in a Kubernetes cluster, data sources and data analysis steps. These custom resources are stored in Kubernetes and can be manipulated in
the same way as regular Kubernetes objects like Pods or Services.
Furthermore, the controller watches for updates to those CRDs and continuously ensure
the desired state. Thus, it automatically creates and delete data analysis containers based on the
custom resource definitions.

%\paragraph{}
\textbf{Data Sources}
This CRD object represents a data source such as a monitored node, virtual machine or pod
in the system. It contains an URL for accessing data and an number of  labels
that describe arbitrary properties of the data source. 
%The labels should contain information like
%the type of monitoring technology used, the monitored component or which system layer is
%monitored. 

%\paragraph{}
\textbf{Analysis Step}
This object defines the analysis workload in the form of a Kubernetes Pod
template. It contains all information necessary to start a data analysis container. The first
parameter of an analysis step is a list of ingest selectors which describe the data sources that
should be consumed and used by the step. These ingest selectors are matched against the labels
of existing data source objects in the system. 
%Together with the key - value pairs which are used
%as selectors, a check property can be used to define different types of match checks, i.e regular expressions or wildcards.
All data sources which match with these ingest selectors are considered appropriate for the analysis step.
Besides the execution instructions, the analysis step definition also contains initial values for all relevant
hyperparameters of the algorithm.

Since we assume that it is not feasible to run every type of analysis steps on lightweight edge devices, we extended 
the operator with a region restriction for analysis steps.
In edge computing environments the infrastructure is often divided into several regions, i.e. 
a public cloud region and several edge regions located anywhere in a smart city \cite{Shi2016}. In addition, some of the edge devices may have additional hardware 
connected - for instance GPUs - which enhances the performance of specific analysis steps.
%The region restriction allows analysis steps to declare that they are not able to run on lightweight edge devices or that they need further hardware.
When an analysis step was defined with a region restriction, the operator only uses nodes which comply with that restriction for 
the scheduling of such analysis steps.

\subsection{Model Repository}
The model repository stores AI models of analysis steps. 
%It can be used in every analysis step to store machine learning models in the repository. 
This is necessary to enable
the warm start of for instance anomaly detection algorithms, which furthers results in a shorter
learning time or even no learning time at all. In addition, models can continuously be updated to adapt to seasonal changes.
We are using the Redis in-memory database\footnote{https://redis.io/} as a model repository since 
it shows promising performance in the IoT context 
\cite{Braulio2018} and further provides modules for AI and edge computing use cases\footnote{\url{https://redislabs.com/redis-enterprise/redis-edge/}}.

%% file: 04-experiments.tex
\section{Performance experiments}
Edge devices are typically restricted in terms of processing power, memory and energy consumption \cite{Wikstrom2014}. 
Therefore, we focus on the overhead of resource usage of for key technology parts inside ZerOps4E:
\begin{itemize}
    \item Resource efficient collecting of metrics on edge devices
    \item Feasibility of deploying multiple anomaly detection models on edge devices
\end{itemize}
In addition, we compare the resource consumption between edge devices and commodity servers, which function as cloud instances to be able to show key differences but also provide guidance for practical usage.
%Since our components were previously only tested
%on cloud environments consisting of commodity servers we have run experiments to evaluate the performance and overhead of the collector component and used anomaly detection algorithms on edge devices. 
%\subsection{Evaluation Setup}
For evaluation, we used the devices depicted in Table \ref{tab:evaldevices}.
The results of the commodity server are compared to the performance on typical edge devices - a Raspberry Pi 3B and Raspberry Pi 4B.

\begin{table}[]
\centering
\caption{Hardware used for the performance evaluation. %The experiments were run on a commodity server, a Raspberry Pi 3B and Raspberry Pi 4B.
}
\label{tab:evaldevices}
\begin{tabular}{|l|c|c|c|}
\hline
                & \textbf{\begin{tabular}[c]{@{}c@{}}Commodity\\ Server\end{tabular}} & \textbf{RPI 3B} & \textbf{RPI 4B} \\ \hline
\textbf{CPU} &
  \begin{tabular}[c]{@{}c@{}}Intel Xeon CPU\\  E3-1230 V2 3.30GHz\end{tabular} &
  \begin{tabular}[c]{@{}c@{}}Cortex-A53\\ 1.4GHz\end{tabular} &
  \begin{tabular}[c]{@{}c@{}}Cortex-A72\\  1.5GHz\end{tabular} \\ \hline
\textbf{Memory} & 16GB                                                                & 1GB             & 2GB             \\ \hline
\textbf{Ethernet} &
  \begin{tabular}[c]{@{}c@{}}10/100/1000 \\ MBit/s\end{tabular} &
  \begin{tabular}[c]{@{}c@{}}10/100/1000 \\ MBit/s\end{tabular} &
  \begin{tabular}[c]{@{}c@{}}10/100/1000\\  MBit/s\end{tabular} \\ \hline
\end{tabular}
\vspace{-1em}
\end{table}

\subsection{Experiment setup}
The platform is envisioned to run on heterogeneous environments in terms of CPU architectures and therefore we
have adjusted our continuous integration (CI) pipeline to compile the components for \textit{amd64}, \textit{arm32v7}, and \textit{arm64v8}.
%Jenkins\footnote{\url{https://www.jenkins.io/}} is used as CI server to cross-compile inside of Docker containers and therefore obviating  the need for otherwise necessary additional ARM hardware.
We are using Docker to provide the build artifacts and create images for each CPU architecture. At the end of each CI pipeline a Docker manifest is created 
which acts as a multi-arch docker image.
%When this image is pulled the Docker daemon on the node will automatically choose the right image for the used
%CPU architecture.
This enables us to use the same deployment scripts for every node in infrastructure, regardless of the CPU architecture.
Therefore our, platform can rapidly be deployed even on heterogeneous infrastructures.

\subsection{Monitoring Overhead}
We evaluated the overhead of the collector with different frequencies while running on the aforementioned edge devices.
On all devices the bitflow-collector was deployed within a docker container and the resource consumption was monitored in order to capture the collector's overhead on the CPU resource consumption. 
We assume that high-frequent collecting of monitoring data provides benefits for data analysis, but we expect limitations within the resource usage overhead, which introduces further noise and limit the number of coexisting service containers. 
The collecting frequency of the monitoring agent was increased by 100ms in an interval of [100ms, 10s].  

\begin{figure}[ht]\centering
\subcaptionbox{Memory overhead of monitoring agent}
    {
    \begin{tikzpicture}[scale=0.7]
            \begin{axis}[
                xlabel={Frequency in ms},
                ylabel={Avg. Mem. [\%]},
                xmin=0, xmax=1000,
                ymin=0, ymax=5,
                %xtick={0,1,2,3,4},
                %ytick={0,200,400,600,800,1000,1200,1400, 1600. 1800, 2000, 2200, 2400 },
                %legend pos=middle west,
                ymajorgrids=true,
                grid style=dashed,width=8cm,height=4.5cm
            ]
            \addplot table[x=ms,y=mem ,col sep=comma] {results/results_average_wally025.csv}; 
            \addlegendentry{Commodity Server}
            \addplot table[x=ms,y=mem ,col sep=comma] {results/results_average_pi4b.csv}; \addlegendentry{Raspberry Pi 4B}
            \addplot table[x=ms,y=mem ,col sep=comma] {results/results_average_pi3b.csv}; \addlegendentry{Raspberry Pi 3B}
            \end{axis}
        \end{tikzpicture}
    }
    \subcaptionbox{CPU overhead of monitoring agent}
        {
        \begin{tikzpicture}[scale=0.7]
            \begin{axis}[
                xlabel={Frequency in ms},
                ylabel={Avg. CPU [\%]},
                xmin=0, xmax=1000,
                ymin=0, %ymax=1400,
                legend pos=north east,
                ymajorgrids=true,
                grid style=dashed,
            ]
            \addplot table[x=ms,y=cpu ,col sep=comma] {results/results_average_wally025.csv}; \addlegendentry{Commodity Server}
            \addplot table[x=ms,y=cpu ,col sep=comma] {results/results_average_pi4b.csv}; \addlegendentry{Raspberry Pi 4B}
            \addplot table[x=ms,y=cpu ,col sep=comma] {results/results_average_pi3b.csv}; \addlegendentry{Raspberry Pi 3B}
            \end{axis}
        \end{tikzpicture}
        }
    \caption{CPU and memory overhead of the the bitflow-collector running on a commodity server and  Raspberry Pi's with collecting frequencies starting from 100ms to 10s.}
    \vspace{-1em}
    \label{collector_overhead}
\end{figure}
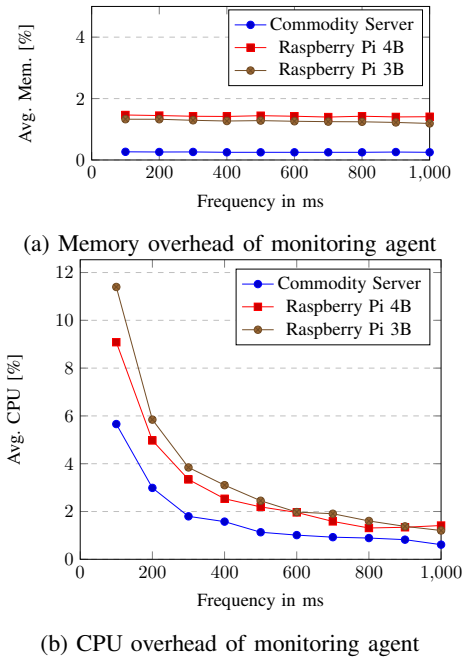

Figure \ref{collector_overhead} provides insights about the memory (a) and (b) CPU relative overhead to the monitored system. 
The left diagram shows that the memory consumption overhead is less than 2\% for the monitoring of any of the hardware components and does not depend on the collection frequency. This is expected as the collected metrics are directly transferred to the local data analysis and not stored in memory. In contrast, the CPU utilization increases for high-frequent collection (e.g. 100ms) compared to collecting metrics every 1 second. This is expected as the resource overhead of collecting increases by the rate of parsing the relevant APIs.
For both, memory as well as CPU utilization, the edge devices consume a higher percentage compared to the commodity server.

The results depicted in Figure \ref{collector_overhead} show, that with a frequency of 500ms the
overhead of the collector amounts to only 2-3\% CPU utilization on
edge devices. Therefore, it is feasible to deploy the bitflow-collector on lightweight edge devices without interfering with running services in the infrastructure, while providing metrics for analysis in high frequency.

\subsection{Performance of Anomaly Detection Algorithms}
The ARIMA, BIRCH, and LSTM-based anomaly detection algorithms were implemented in Java and integrated into the Bitflow framework.
We are using Docker as execution platform, which allows us to limit the resources of running processes through cgroups. For comparison, we performed the following experiment on a commodity server, a Raspberry Pi 3B and a Raspberry Pi 4B.

For each experiment, a data set of 10,000 sample with 28 monitoring metrics was read from a file and  the respective algorithms were applied. While execution, we measured the average processing time per sample, and their standard deviation. In case of the commodity server, we started with no resource limitation at all
(8 virtual CPUs) and decreased the allocated vCPUs by 0.1 for each following run until reaching the minimum assignable
resources at 0.1 vCPUs. 

For the Raspberry Pis, we compiled a custom Raspbian Linux kernel since the default kernel does not support \textit{cpu.cfs\_period\_us} and 
\textit{cpu.cfs\_quota\_us} cgroups used by the Docker engine for resource limitations.
Afterwards, we ran the same experiments starting with 4 vCPUs and decreasing it again by 0.1 vCPU until 0.1 vCPUs are reached.

\begin{figure*}[ht]
    \centering
    \subcaptionbox{Commodity Server}
    {
        \begin{tikzpicture}[scale=0.65]
            \begin{axis}[
                xlabel={CPUs},
                ylabel={ms/sample},
                xmin=0, xmax=8,
                ymin=0, ymax=1400,
                xtick={0,1,2,3,4,5,6,7,8},
                ytick={0,200,400,600,800,1000,1200,1400},
                legend pos=north east,
                ymajorgrids=true,
                grid style=dashed,
            ]       
            \addplot table[x=cpu,y=CABIRCH,col sep=comma] {results/laptop.csv}; \addlegendentry{BIRCH}
            \addplot table[x=cpu,y=LSTM,col sep=comma] {results/laptop.csv}; \addlegendentry{LSTM}
            \addplot table[x=cpu,y=ARIMA,col sep=comma] {results/laptop.csv}; \addlegendentry{ARIMA}
            \end{axis}
        \end{tikzpicture}
    }
    \subcaptionbox{Raspberry Pi 4B}
        {
        \begin{tikzpicture}[scale=0.65]
            \begin{axis}[
                xlabel={CPUs},
                ylabel={s/sample},
                xmin=0, xmax=4,
                ymin=0, %ymax=20,
                xtick={0,1,2,3,4},
                legend pos=north east,
                ymajorgrids=true,
                grid style=dashed,
            ]
            \addplot table[x=cpu,y expr=\thisrow{CABIRCH} * 0.001 ,col sep=comma] {results/rpi4.csv}; \addlegendentry{BIRCH}
            \addplot table[x=cpu,y expr=\thisrow{LSTM} * 0.001 ,col sep=comma] {results/rpi4.csv}; \addlegendentry{LSTM}
            \addplot table[x=cpu,y expr=\thisrow{ARIMA} * 0.001 ,col sep=comma] {results/rpi4.csv}; \addlegendentry{ARIMA}

            \end{axis}
        \end{tikzpicture}
        }
        \subcaptionbox{\footnotesize Raspberry Pi 3B}
        {
        \begin{tikzpicture}[scale=0.65]
            \begin{axis}[
                xlabel={CPUs},
                ylabel={s/sample},
                xmin=0, xmax=4,
                ymin=0, %ymax=20,
                xtick={0,1,2,3,4},
                legend pos=north east,
                ymajorgrids=true,
                grid style=dashed,
            ]
            \addplot table[x=cpu,y expr=\thisrow{CABIRCH} * 0.001 ,col sep=comma] {results/rpi3b.csv}; \addlegendentry{BIRCH}
            \addplot table[x=cpu,y expr=\thisrow{LSTM} * 0.001 ,col sep=comma] {results/rpi3b.csv}; \addlegendentry{LSTM}
            \addplot table[x=cpu,y expr=\thisrow{ARIMA} * 0.001 ,col sep=comma] {results/rpi3b.csv}; \addlegendentry{ARIMA}

            \end{axis}
        \end{tikzpicture}
        }
    \caption{Illustrated processing time per sample under varying resource limitations. Note that the processing time for the Raspberry Pi is depicted in seconds, whereas for the commodity server it is depicted in milliseconds.}
    \vspace{-1.5em}
    \label{aa_experiments}
\end{figure*}
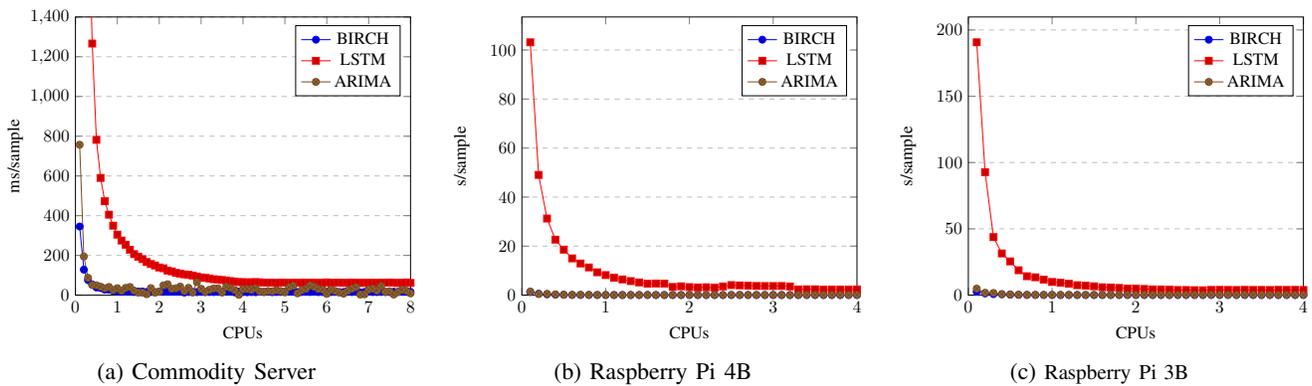

Figure \ref{aa_experiments} shows the results of the evaluation for the three devices: commodity server (a), Raspberry Pi 4B (b), Raspberry Pi 3B (c). All three plots show the same general behavior of exponential increase when limiting the CPU resources towards zero. This is expected as, by limiting the CPU utilization, the algorithms have less computational resources, while still running the same processing of samples. Consequently, the runtime per sample grows. 
The applied algorithms on the commodity server depict the shortest computation times compared to the edge devices as the commodity server utilizes a more powerful CPU chip.
The results show that BIRCH and ARIMA perform more efficient than LSTM in terms of time used per sample. This is due to the complexity of the applied algorithms: The anomaly detection algorithms are continuously trained and the deep learning models suffer from a higher computational complexity due to backpropagation while training. Consequently, the deep learning models are expected to have a higher computation time.

\begin{figure*}[ht]
\subcaptionbox{CPU overhead (monitoring \& ARIMA)}
    {
    \begin{tikzpicture}[scale=0.65]
            \begin{axis}[
                ybar,
                xlabel={Frequency in ms},
                ylabel={Average CPU utilization in \%},
                xmin=100,% xmax=1100,
                ymin=0, %ymax=1400,
                xtick={200,400,600,800, 1000},
                legend pos=north east,
                ymajorgrids=true,
                grid style=dashed,
            ]
            \addplot table[x=ms,y expr=\thisrow{cpu}+\thisrow{adcpuarima} ,col sep=comma] {results/results_average_wally025_neu.csv}; \addlegendentry{Commodity Server}
            \addplot table[x=ms,y expr=\thisrow{cpu}+\thisrow{adcpuarima} ,col sep=comma] {results/results_average_pi4b_neu.csv}; \addlegendentry{Raspberry Pi 4B}
            \addplot table[x=ms,y expr=\thisrow{cpu}+\thisrow{adcpuarima} ,col sep=comma] {results/results_average_pi3b_neu.csv}; \addlegendentry{Raspberry Pi 3B}
            \end{axis}
        \end{tikzpicture}
    }
    \subcaptionbox{CPU overhead (monitoring \& BIRCH)}
        {
        \begin{tikzpicture}[scale=0.65]
            \begin{axis}[
                ybar,
                xlabel={Frequency in ms},
                ylabel={Average CPU utilization in \%},
                xmin=100,% xmax=1000,
                ymin=0, %ymax=1400,
                xtick={200,400,600,800, 1000},
                legend pos=north east,
                ymajorgrids=true,
                grid style=dashed,
            ]
            \addplot table[x=ms,y expr=\thisrow{cpu}+\thisrow{adcpucabirch} ,col sep=comma] {results/results_average_wally025_neu.csv}; \addlegendentry{Commodity Server}
            \addplot table[x=ms,y expr=\thisrow{cpu}+\thisrow{adcpucabirch} ,col sep=comma] {results/results_average_pi4b_neu.csv}; \addlegendentry{Raspberry Pi 4B}
            \addplot table[x=ms,y expr=\thisrow{cpu}+\thisrow{adcpucabirch} ,col sep=comma] {results/results_average_pi3b_neu.csv}; \addlegendentry{Raspberry Pi 3B}
            \end{axis}
        \end{tikzpicture}
        }
        \subcaptionbox{CPU overhead (monitoring \& LSTM)}
        {
        \begin{tikzpicture}[scale=0.65]
            \begin{axis}[
                ybar,
                xlabel={Frequency in ms},
                ylabel={Average CPU utilization in \%},
                xmin=100, %xmax=1000,
                ymin=0, ymax=150,
                legend pos=north east,
                ymajorgrids=true,
                grid style=dashed,
            ]
            \addplot table[x=ms,y expr=\thisrow{cpu}+\thisrow{adcpulstm} ,col sep=comma] {results/results_average_wally025_neu.csv}; \addlegendentry{Commodity Server}
            \addplot table[x=ms,y expr=\thisrow{cpu}+\thisrow{adcpulstm} ,col sep=comma] {results/results_average_pi4b_neu.csv}; \addlegendentry{Raspberry Pi 4B}
            \addplot table[x=ms,y expr=\thisrow{cpu}+\thisrow{adcpulstm} ,col sep=comma] {results/results_average_pi3b_neu.csv}; \addlegendentry{Raspberry Pi 3B}
            \end{axis}
        \end{tikzpicture}
        }
    \caption{CPU utilization of the monitoring agent and anomaly detection algorithms  for monitoring frequencies from 200ms to 1s during just-in-time processing of samples.}
    \vspace{-1.5em}
    \label{overall_overhead}
\end{figure*}
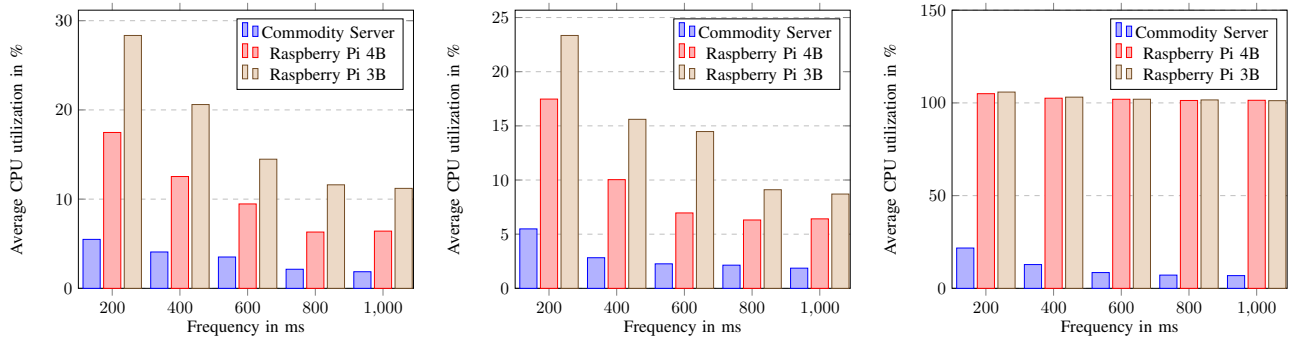

The results show, that BIRCH and ARIMA are computation efficient on both, the edge device and commodity server. In both cases, we expect that for vCPU limitations with maximum of 0.6, the algorithms are capable to be applied in an AIOps use-case as the monitoring rates are expected to be 500ms or higher. This is also depicted in Figure \ref{overall_overhead} which shows the combined CPU overhead of the monitoring agent and the anomaly detection algorithms when samples are processed just-in-time, meaning at the same speed new samples are provided in the data stream.

Applying deep learning models with continuous training may not suite the needs of real-time processing on edge devices as the computation time is likely to be higher than the monitoring rate. 
%Thus, it is higher than a monitoring data rate of e.g. 500ms. 
For such techniques, the anomaly detection should be moved to the cloud. Nevertheless, we showed that deep learning approaches are capable of detecting in high quality \cite{Schmidt2018}. For complex behavior pattern detection, such approaches might be applied for latency unrelated edge use cases, while real-time responsible devices should utilize BIRCH or ARIMA.E

In conclusion, the limitation of resources can be used to establish SLAs to the self-healing capabilities on edge devices for applying AI algorithms on edge devices.
Furthermore, anomaly detection based on BIRCH or ARIMA are both feasible to be applied on edge devices, while LSTM based anomaly detection should be considered to run in the cloud for potential higher qualitative results. 
In future, the combination of quick responses at the edge could be aggregated with higher qualitative results from cloud services. 

%% file: 07-summary.tex
\section{Conclusion}
The paper described the framework ZerOps4E in order to apply self-healing capabilities to distributed heterogeneous infrastructures.
The architecture of ZerOps4E includes several different key components including resource monitoring and analysis steps, which are scheduled distributed and are capable to run on small powered edge devices. 
We showed the feasibility of deploying high-frequent resource monitoring agents directly on edge devices as they can be applied resource-efficient.
Besides simple aggregation and filtering methods applied on the monitored data inside edge devices, more sophisticated techniques might be applied like anomaly detection. For such methods, we showed that there exist multiple anomaly detection solutions, which are applicable for edge devices, while others like deep-learning models might be not applicable. For such cases, one has to decide on further properties like quality of anomaly detection approaches and latency guarantees between possible execution places.